\documentclass[%
 aip,
 amsmath,amssymb,
 reprint,%
]{revtex4-1}

\usepackage{graphicx}
\usepackage{dcolumn}
\usepackage{bm}

\usepackage[utf8]{inputenc}
\usepackage[T1]{fontenc}
\usepackage{mathptmx}

\begin{document}

\preprint{AIP/123-QED}

\title{FFT-based free space Poisson solvers: why Vico-Greengard-Ferrando should replace Hockney-Eastwood}
\author{Junyi Zou, Eugenia Kim, and Antoine J. Cerfon}

\affiliation{Courant Institute of Mathematical Sciences, New York University}%

\date{\today}

\begin{abstract}
    Many problems in beam physics and plasma physics require the solution of Poisson's equation with free-space boundary conditions. The algorithm proposed by Hockney and Eastwood is a popular scheme to solve this problem numerically, used by many cutting-edge codes, because of its speed and its simplicity. However, the potential and its gradient obtained with this method have low accuracy, and the numerical error converges slowly with the number of grid points. We demonstrate that the closely related algorithm recently proposed by Vico, Greengard, and Ferrando is just as easy to implement, just as efficient computationally, and has much higher accuracy, with a rapidly converging error. 
\end{abstract}

\maketitle


\section{Motivation}\label{sec:Intro}
In many simulations in beam physics and plasma physics, one needs to solve Poisson's equation for the electrostatic potential $\phi$ due to the charge density $\rho$
\begin{equation}\label{eq:Poisson}
    \Delta\phi=-\rho
\end{equation}
with free-space boundary conditions, such that the electric field $\mathbf{E}=-\nabla\phi$ vanishes infinitely far from the charge distribution \cite{Daughton2006fully,Qiang2006Quasistatic,Yang2010beam,CerfonPRSTAB2013,CerfonPRL2016, Guadagni2017fast}. This is the case when the physical system is such that the charge distribution is indeed in free space. 
Following standard potential theory, the solution to this problem can also serve as the particular solution entering in the construction of the full solution of Poisson's equation on general domains and with more general boundary conditions \cite{Askham2017adaptive,Fryklund2018partition}.

The natural solution to Eq.\eqref{eq:Poisson} with free-space boundary conditions is to write 
\begin{equation}\label{eq:convolution}
    \phi(\mathbf{r})=\int_{\mathbb{R}^{d}}\rho(\mathbf{r}')G(\mathbf{r}-\mathbf{r}')d\mathbf{r}'
\end{equation}
where $G(\mathbf{r})=-\frac{1}{2\pi}\ln r$ for the two-dimensional problem, and $G(\mathbf{r})=\frac{1}{4\pi r}$ for the three-dimensional problem, and $r=|\mathbf{r}|$. In most applications of physical interest, the charge distribution $\rho$ has compact support within the computational domain. There are two well-known difficulties with the numerical computation of \eqref{eq:convolution}. First, because of the long range of the Green's functions, fast algorithms are necessary for the efficient computation of the convolution. Second, because of the singularity of the Green's functions at $\mathbf{r}=\mathbf{r}'$, special quadrature schemes must be designed to obtain high accuracy. In the last two decades, numerical schemes have been developed to address both difficulties in an efficient manner, even for cases where the charge distribution $\rho$ is highly inhomogeneous and requires adaptive discretization \cite{Ethridge2001,Mccorquodale2005scalable,langston2011,Malhotra2015pvfmm}. 

The purpose of this letter is not to revisit these well-established algorithms. Instead, our goal is to focus on the common situation in beam physics and plasma physics where \eqref{eq:Poisson} is solved in a Cartesian coordinate system, and where the grid points are uniformly spaced. In such cases, the Fast-Fourier transform (FFT) is an efficient tool to address the first difficulty. Pioneering in that direction, Hockney and Eastwood \cite{hockney1988computer} have shown that by appropriately zero padding the charge distribution $\rho$, the aperiodic convolution \eqref{eq:convolution} could be computed ``exactly", i.e. to within errors due to the discretization and the quadrature scheme, with FFTs. Specifially, according to the convolution theorem, the convolution \eqref{eq:convolution} is transformed to a product of functions in the Fourier domain. Thus, discretizing the charge distribution on the uniform grid, zero padding this discretized distribution in each spatial direction on an interval of the same length as the original domain, and defining a periodic Green's function $\tilde{G}$ on that extended domain, they compute the electrostatic potential according to:
\begin{equation}
    \phi=h_{x}h_{y}h_{z}FFT^{-1}\{FFT\{\tilde{\rho}\}FFT\{\tilde{G}\}\}
\end{equation}
where $h_{x}$, $h_{y}$, and $h_{z}$ represent the grid spacing in each Cartesian dimension, $\tilde{\rho}$ is the padded charge distribution on the extended domain, $FFT\{\cdot\}$ represents an FFT in all spatial dimensions, and $FFT^{-1}\{\cdot\}$ represents an inverse FFT in all spectral dimensions. The numerical solver based on this approach is remarkably simple to implement, and efficient, since the computational complexity scales like $\mathcal{O}((2N)^d(\log 2N)^d)$, where $N$ is the number of grid points in each direction for the original computational domain, before padding. As a result, it is widely used, including in high-performance codes for accelerator science and for plasma physics applications \cite{Budiardja2011parallel,Shishlo2015particle,Adelmann2019opal}.

However, a weakness of this scheme is the slow decay of the numerical error with increasing grid size: if the mesh spacing is $h$ in each spatial direction, the scheme is a second order scheme, with the error decaying like $h^2$. This is because the method relies on the standard trapezoidal rule for the discretization of \eqref{eq:convolution}, together with a regularization of the Green's function, chosen such that $G(\mathbf{0})=1$, which is not designed to achieve high accuracy \cite{Rasmussen2011particle}. Alternate simple regularizations have been suggested\cite{Chatelain2010,Budiardja2011parallel}, which do not improve the order of convergence of the method, as discussed in \cite{Rasmussen2011particle,Hejlesen2019non}.

In contrast, Vico, Greengard, and Ferrando  have recently proposed a closely related scheme, which is as easy to implement, as efficient computationally, but has a high order of convergence, giving the potential and the electric field with high accuracy for a modest number of grid points \cite{Vico2016fast}. The purpose of this letter is to encourage beam and plasma physicists to switch from Hockney \& Eastwood to this scheme, by briefly explaining the numerical method it is based on, showing that its implementation is indeed as simple, and providing test cases demonstrating the superiority of the method.

The remainder of this article is structured as follows. In Section \ref{sec:review}, we give a brief overview of the algorithm proposed by Vico, Greengard, and Ferrando. In Section \ref{sec:compare}, we compare the accuracy of the algorithm of Hockney and Eastwood, and several of its common variants, with the accuracy of the algorithm by Vico \textit{et al.}. We then conclude with a brief summary.

\section{Brief description of the scheme by Vico, Greengard, and Ferrando}\label{sec:review}

Without loss of generality, let us assume that the computational domain is the unit box $D=[-0.5,  0.5]^d$. As we discussed previously, we operate under the assumption that $\rho$ is compactly supported in $D$. For both $d=2$ and $d=3$, and for $L>0$, we define the following windowed Green's functions:
\begin{equation}\label{eq:windowed}
    G^{L}(\mathbf{r})=G(\mathbf{r})\mathbf{1}_{I}\left(\frac{r}{2L}\right)
\end{equation}
where $\mathbf{1}_{I}$ is the indicator function for the interval $I=[-0.5,0.5]$, such that $\mathbf{1}_{I}(\mathbf{x})=1$ if $\mathbf{x}\in I$, and $\mathbf{1}_{I}(\mathbf{x})=0$ otherwise. Note that in the computational domain, the maximum distance between a source at $\mathbf{r}'$ and a target at $\mathbf{r}$ is $\sqrt{d}$. The first key observation by Vico \textit{et al.} is that by setting $L>\sqrt{d}$, we have for all $\mathbf{r}$ in $D$:
\begin{equation}\label{eq:PoissonWindowed}
    \phi(\mathbf{r})=\int_{D}\rho(\mathbf{r}')G(\mathbf{r}-\mathbf{r}')d\mathbf{r}'=\int_{D}\rho(\mathbf{r}')G^{L}(\mathbf{r}-\mathbf{r}')d\mathbf{r}'
\end{equation}
If we define the Fourier transform $\mathcal{F}\{f\}$ of a function $f$, and the inverse Fourier transform $\mathcal{F}^{-1}\{\hat{f}\}$ of a function $\hat{f}$, the electrostatic potential in the computational domain $D$ can thus be computed according to:
\begin{equation}\label{eq:PoissonFourier}
    \phi=\mathcal{F}^{-1}\{\mathcal{F}\{\rho\}\mathcal{F}\{G^L\}\}
\end{equation}
The second key observation of Vico \textit{et al.} is that $\mathcal{F}\{G^L\}$ can be computed analytically, and that it is regular at $\mathbf{k}=\mathbf{0}$, where $\mathbf{k}$ is the Fourier wavevector, unlike the original Green's function, whose Fourier transform is $\mathcal{F}\{G\}=\frac{1}{|\mathbf{k}|^2}$. We have
\begin{align}
  &\mathcal{F}\{G^{L}\}=\frac{1-J_{0}(L|\mathbf{k}|)}{|\mathbf{k}|^2}-\frac{L\ln L J_{1}(L|\mathbf{k}|) }{|\mathbf{k}|}\qquad (\mbox{in 2D})\label{GLFourier2D}\\
  &\mathcal{F}\{G^{L}\}=2\left(\frac{\sin\left(\frac{L|k|}2\right)}{|k|}\right)^{2}\qquad (\mbox{in 3D})\label{GLFourier3D}
\end{align}
We can therefore write the following explicit formula for the solution to Poisson's equation with free-space boundary conditions in two dimensions
\begin{equation}
    \phi(\mathbf{r})=\frac{1}{4\pi^2}\iint_{\mathbb{R}^2}e^{i\mathbf{k}\cdot \mathbf{r}}\mathcal{F}\{\rho\}(\mathbf{k})\left[\frac{1-J_{0}(L|\mathbf{k}|)}{|\mathbf{k}|^2}-\frac{L\ln L J_{1}(L|\mathbf{k}|) }{|\mathbf{k}|}\right]d\mathbf{k}
\end{equation}
and an analogous formula for three dimensions. Remarkably, the term in the square brackets in the integrand is an analytic function, which is a direct consequence of the Paley-Wiener theorem. It is also a nonincreasing function of $\mathbf{k}$. Discretization of the integral by the trapezoidal rule will thus allow the computation of $\phi$ with high accuracy, with the order of convergence of the scheme determined by the rate of decay of $\mathcal{F}\{\rho\}(\mathbf{k})$, and spectral accuracy achieved for sufficiently smooth $\rho(\mathbf{r})$ \cite{Trefethen2000spectral,Trefethen2014trapezoidal}, as commonly encountered in practical applications. The algorithm can therefore rely solely on standard FFTs for the fast evaluation of the aperiodic convolution, as for Hockney and Eastwood. 

Because of the oscillatory nature of $\mathcal{F}\{G^L\}$, the charge distribution must in principle be extended by zero-padding to a grid of size $(4N)^{d}$, and the computations done on this extended grid, in order to compute the integral without aliasing error. This would seem to be a significant drawback as compared to the computations on a grid of size $(2N)^d$ in the Hockney and Eastwood algorithm. In practice, this is however not an issue, because Vico \textit{et al.} show that for a given computational grid and a given Green's function, all Poisson solves can be done on a grid of size $(2N)^d$ after an initial precomputation step on a grid of size $(4N)^{d}$. Specifically, one computes the inverse FFT of $\mathcal{F}\{G^L\}$ on a grid of size $(4N)^{d}$, and the result restricted to the appropriate grid of size $(2N)^d$ then serves as the numerical Green's function for all the other computations, on a grid of size $(2N)^{d}$. This is crucial, since in most applications, Poisson's equation must be solved for a new charge distribution at each time step, on a fixed grid. The precomputation step is then a negligible part of the total computational cost. To summarize, the only difference in terms of numerical implementation between the Hockney \& Eastwood method and the method by Vico \textit{et al.} is that with the former, the Green's function is given analytically, whereas with the latter, it must be precomputed via an inverse FFT of \eqref{GLFourier2D} or \eqref{GLFourier3D} on a domain of size $(4N)^d$. MATLAB versions of our codes can be downloaded on https://www.math.nyu.edu/~cerfon/codes.html for readers further interested in the details of the (easy) implementation.

\section{Numerical comparison of the solvers}\label{sec:compare}
In this section, we compare the accuracy and convergence of the schemes by Hockney \& Eastwood and by Vico \textit{et al.}. For conciseness, we only show one two-dimensional example, and one three-dimensional example, which have radially symmetric and spherically symmetric charge distributions. We obtain similar results for distributions without such symmetry.

\subsection{Two-dimensional example}
We consider here the Gaussian charge density
\begin{equation}
    \rho=\frac{1}{2\pi\sigma^2}\exp\bigg(-\frac{x^2+y^2}{2\sigma^2}\bigg)
\end{equation}
with $\sigma=0.05$, and the computational domain is $[-0.5, 0.5]^2$.

The analytic solution $\phi^*$ for the potential is given by:
\begin{equation}
    \phi^*(x,y)=-\frac{1}{4\pi}\bigg[\text{Ei}\bigg(\frac{x^2+y^2}{2\sigma^2}\bigg)+\log(x^2+y^2)\bigg]
\end{equation}
where $\text{Ei}$ is the exponential integral function. Let $h=1/N$ and $\phi_{i,j}$ be the numerical solution at grid point $[-0.5+ih,-0.5+hj]$. We compute the relative error $e_{i,j}$ as:
\begin{equation}
    e_{i,j}=\bigg|\frac{\phi_{i,j}-\phi^*(-0.5+ih,-0.5+jh)}{\phi^*(-0.5+ih,-0.5+jh)}\bigg|
\end{equation}

\begin{figure}
    \centering
    \includegraphics[width=\linewidth]{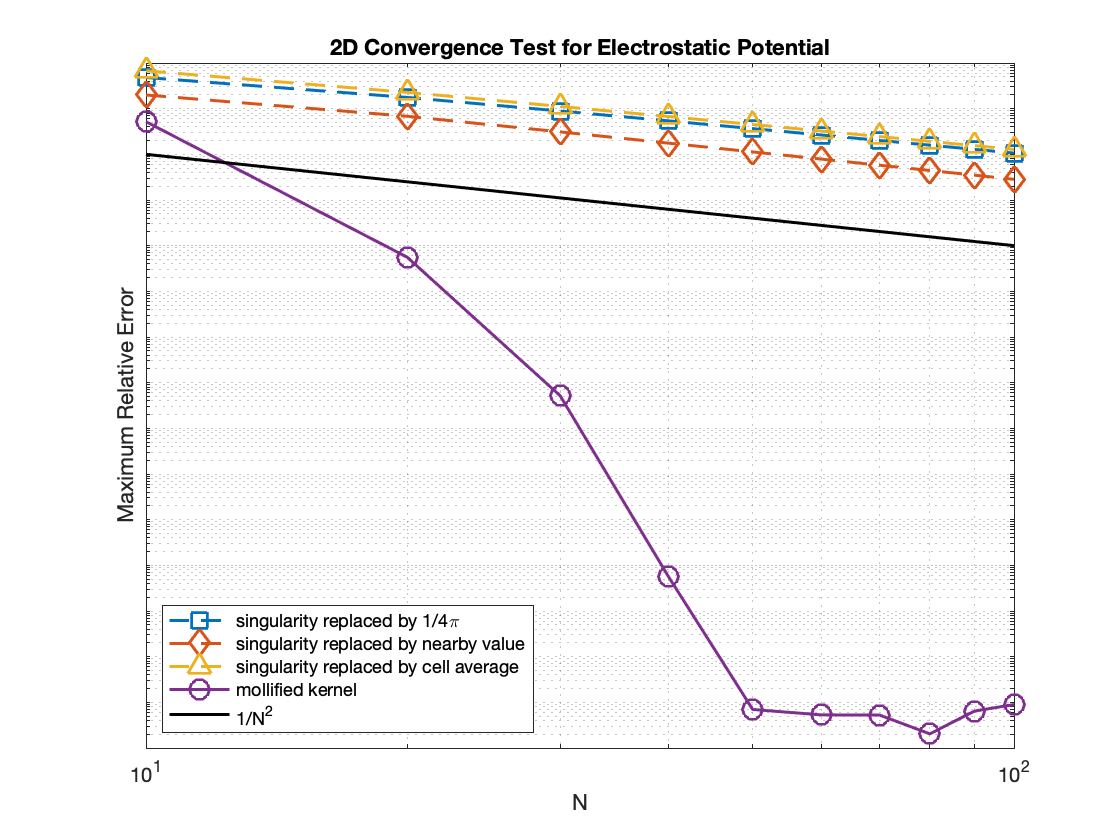}
    \caption{Numerical error for the electrostatic potential in our two-dimensional test, for the different algorithms discussed in this article.}
    \label{fig:2dpot}
\end{figure}

Figure \ref{fig:2dpot} shows the maximal relative error of $\phi$ over all grid points for the method by Vico \textit{et al.} (indicated by "mollified kernel"), the Hockney and Eastwood method (indicated by "singularity replaced by $1/2\pi$" and two variants of Hockney and Eastwood (indicated by "nearby value"\cite{Budiardja2011parallel} and "cell average"\cite{Chatelain2010} respectively), for $N$ ranging from $10$ to $100$. 

To further illustrate our point, we also compute the electric field $\mathbf{E}$. For the FFT-based schemes we consider here, a natural approach is to compute $\mathbf{E}$ through Fourier differentiation (indicated by "Fourier"), i.e. multiplication by the wavevector in Fourier space before computing the inverse FFT. However, centered differences are still commonly used in combination with the Hockney \& Eastwood algorithm, so we also implemented a finite difference approach (indicated by "Finite Diff") for an Hockney \& Eastwood case. Note that since the analytic solution of the electric field is equal to zero at certain grid points, we used the absolute error $e_{abs}$ for this calculation:
\begin{equation}
    e^{abs}_{i,j}=|E_{i,j}-E^*(-0.5+ih,-0.5+jh)|
\end{equation}
\begin{figure}
    \centering
    \includegraphics[width=\linewidth]{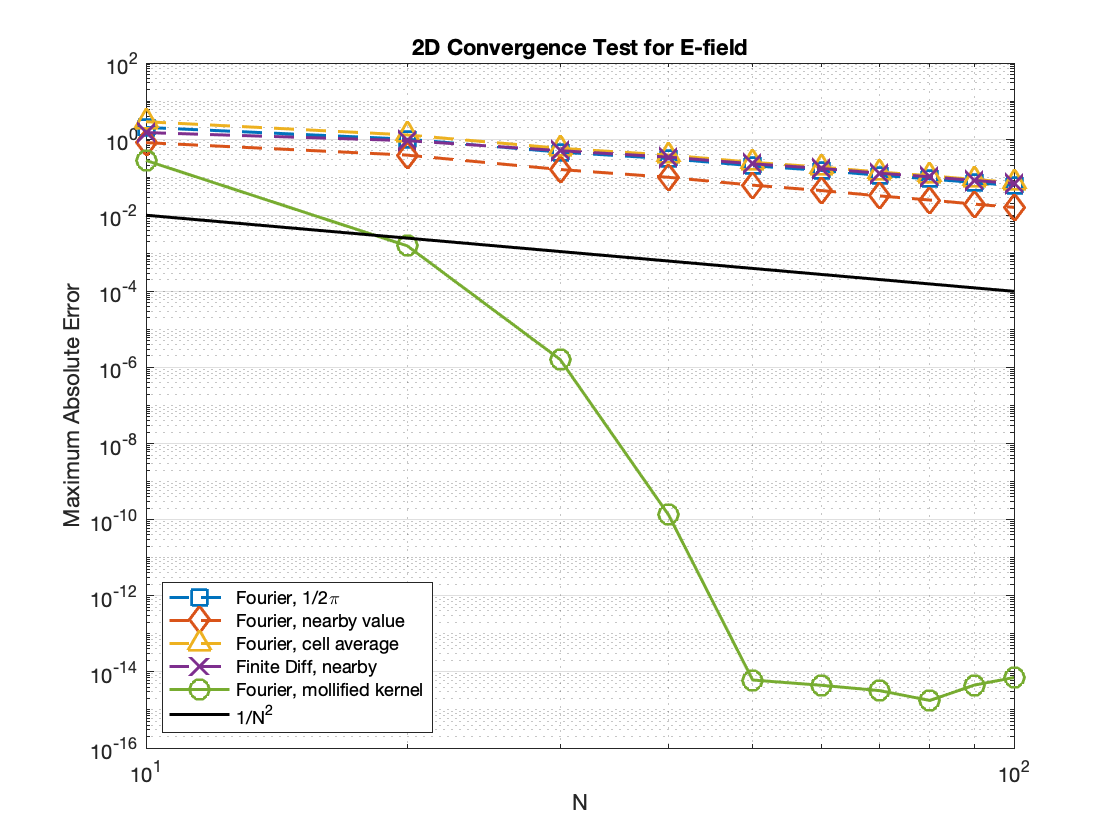}
    \caption{Numerical error for the x-component of the electric field in our two-dimensional test, for the different algorithms discussed in this article. By symmetry, the convergence properties for the y-component are identical.}
    \label{fig:2defield}
\end{figure}
Figure \ref{fig:2defield} shows the maximal absolute error for  $E_x=\frac{\partial\phi}{\partial x}$, for $N$ ranging from $10$ to $100$. We omit the convergence plots for $E_y$ due to the symmetry of the charge density. Figures \ref{fig:2dpot} and \ref{fig:2defield} demonstrate the spectral convergence of the scheme by Vico \textit{et al.}, the second order convergence of the Hockney \& Eastwood algorithms, and that Vico \textit{et al.} leads to much higher accuracy, even for a modest number of grid points.

\subsection{Three-dimensional example}
\begin{figure}
    \centering
    \includegraphics[width=\linewidth]{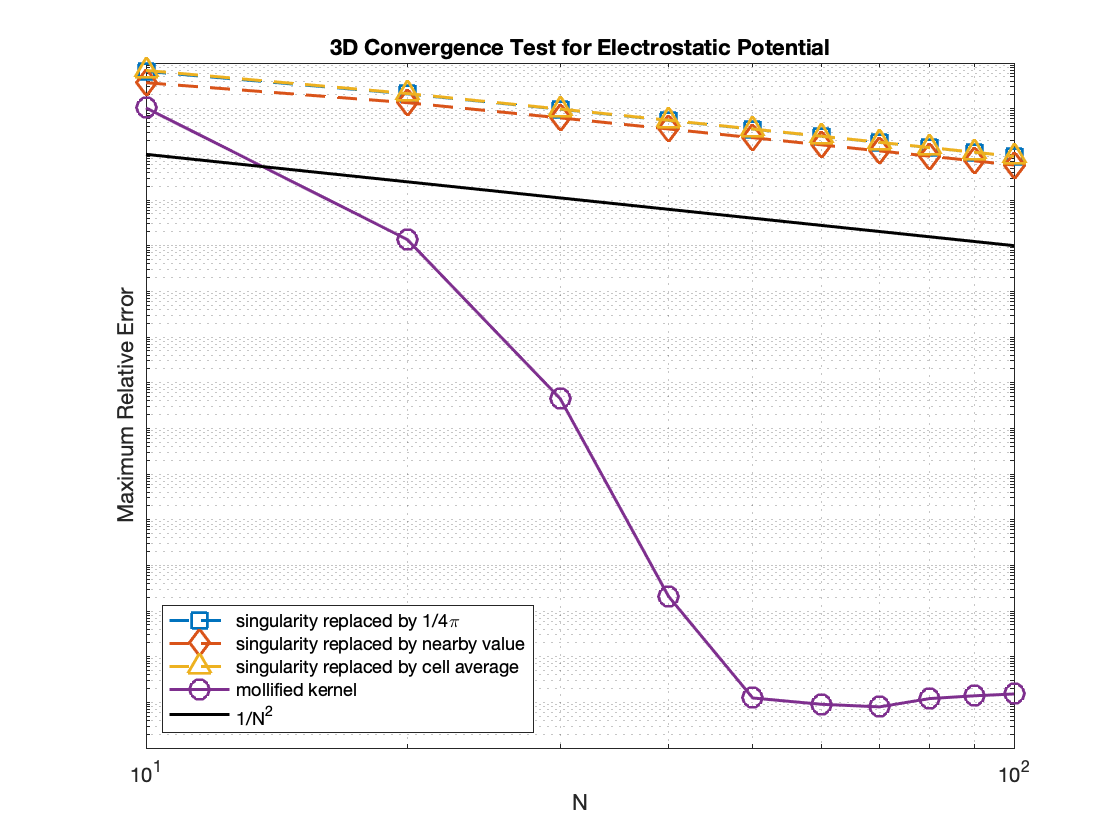}
    \caption{Numerical error for the electrostatic potential in our three-dimensional test, for the different algorithms discussed in this article.}
    \label{fig:3dpot}
\end{figure}

\begin{figure}
    \centering
    \includegraphics[width=\linewidth]{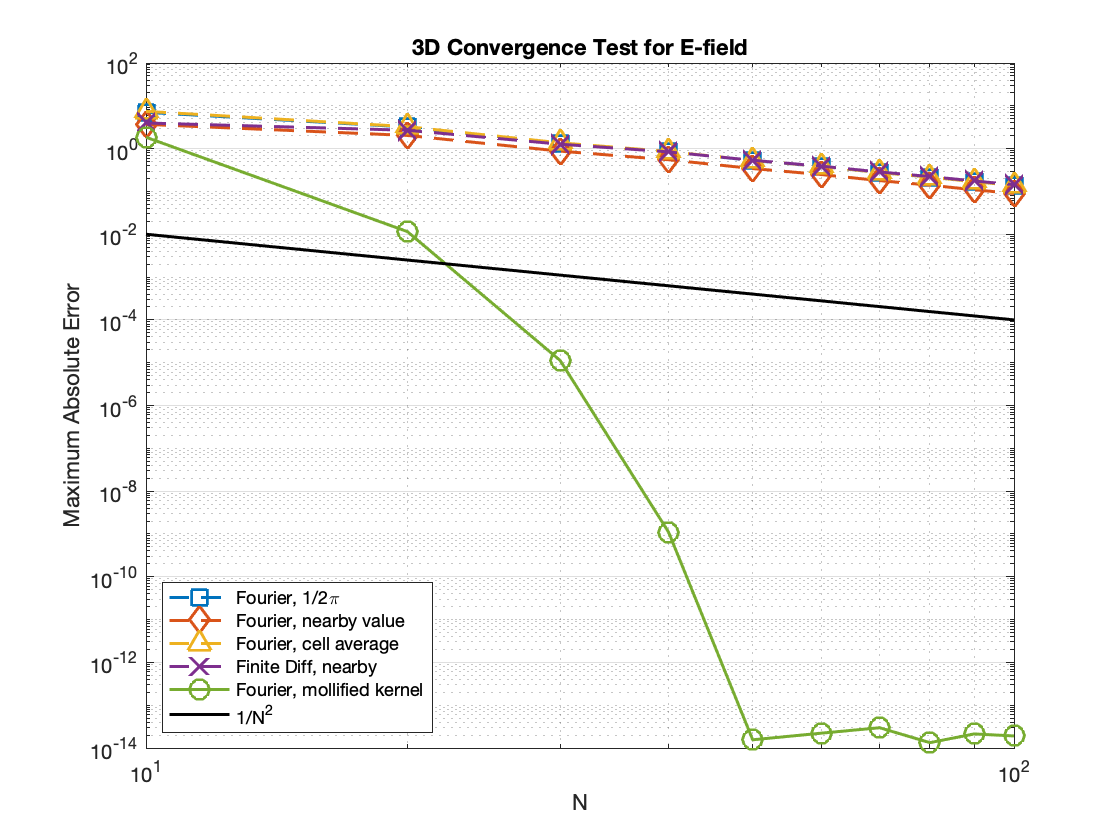}
    \caption{Numerical error for the x-component of the electric field in our three-dimensional test, for the different algorithms discussed in this article. By symmetry, the convergence properties for the y/z-components are identical.}
    \label{fig:3defield}
\end{figure}

We repeat the numerical experiment for the 3-dimensional domain $[-0.5,0.5]^3$, for the following source function and exact solution:
\begin{align}
    \rho&=\frac{1}{(2\pi)^{3/2}\sigma^3}\exp\bigg(-\frac{x^2+y^2+z^2}{2\sigma^2}\bigg)\\
    \phi^*(x,y,z)&=\frac{1}{4\pi\sqrt{x^2+y^2+z^2}}\text{erf}\bigg(\frac{\sqrt{x^2+y^2+z^2}}{\sqrt{2}\sigma}\bigg)
\end{align}
with $\sigma=0.05$. Figure \ref{fig:3dpot} and \ref{fig:3defield} show results similar to the two-dimensional case, again for $N$ ranging from $10$ to $100$.

\section{Concluding remarks}

For the smooth source distributions commonly encountered in beam and plasma physics, the FFT-based free space Poisson solver by Vico, Greengard, and Ferrando is superior to the method by Hockney and Eastwood, with much higher convergence rate and accuracy. Since both algorithms have a comparable computational complexity, and are as simple to implement, our results may encourage users of the popular Hockney \& Eastwood method to switch to the scheme by Vico \textit{et al.}.


We note that Poisson's equation with free-space boundary conditions is also a key element in many astrophysical simulations \cite{BertschingerAstro1998,Colombi_alard_2017}, for the computation of the self-consistent gravitational force. Many solvers rely on the Hockney-Eastwood algorithm \cite{Tanaka2017multidimensional} and other low-order accurate FFT accelerated algorithms \cite{Passy2014adaptive,Moon2019fast,Krasnopolsky_2021}; they would benefit from replacing these schemes with the Vico-Greengard-Ferrando scheme.

Finally, we stress that when the source distribution is not smooth and is highly inhomogeneous, neither of the two FFT-based algorithms described here gives solutions with high accuracy. When small errors are required for these cases, advanced adaptive algorithms should be favored\cite{Ethridge2001,Mccorquodale2005scalable,langston2011,Malhotra2015pvfmm}.

\begin{acknowledgments}
We would like to thank L. Greengard and F. Vico for their insightful comments. We also acknowledge support from the United States National Science Foundation under Grant No. PHY-1820852
\end{acknowledgments}

\section*{Data availability}

The data that support the findings of this study are available
from the corresponding author upon reasonable request.

\bibliography{PoissonSolve}

\providecommand{\noopsort}[1]{}\providecommand{\singleletter}[1]{#1}%
\begin{thebibliography}{28}%
\makeatletter
\providecommand \@ifxundefined [1]{%
 \@ifx{#1\undefined}
}%
\providecommand \@ifnum [1]{%
 \ifnum #1\expandafter \@firstoftwo
 \else \expandafter \@secondoftwo
 \fi
}%
\providecommand \@ifx [1]{%
 \ifx #1\expandafter \@firstoftwo
 \else \expandafter \@secondoftwo
 \fi
}%
\providecommand \natexlab [1]{#1}%
\providecommand \enquote  [1]{``#1''}%
\providecommand \bibnamefont  [1]{#1}%
\providecommand \bibfnamefont [1]{#1}%
\providecommand \citenamefont [1]{#1}%
\providecommand \href@noop [0]{\@secondoftwo}%
\providecommand \href [0]{\begingroup \@sanitize@url \@href}%
\providecommand \@href[1]{\@@startlink{#1}\@@href}%
\providecommand \@@href[1]{\endgroup#1\@@endlink}%
\providecommand \@sanitize@url [0]{\catcode `\\12\catcode `\$12\catcode
  `\&12\catcode `\#12\catcode `\^12\catcode `\_12\catcode `\%12\relax}%
\providecommand \@@startlink[1]{}%
\providecommand \@@endlink[0]{}%
\providecommand \url  [0]{\begingroup\@sanitize@url \@url }%
\providecommand \@url [1]{\endgroup\@href {#1}{\urlprefix }}%
\providecommand \urlprefix  [0]{URL }%
\providecommand \Eprint [0]{\href }%
\providecommand \doibase [0]{http://dx.doi.org/}%
\providecommand \selectlanguage [0]{\@gobble}%
\providecommand \bibinfo  [0]{\@secondoftwo}%
\providecommand \bibfield  [0]{\@secondoftwo}%
\providecommand \translation [1]{[#1]}%
\providecommand \BibitemOpen [0]{}%
\providecommand \bibitemStop [0]{}%
\providecommand \bibitemNoStop [0]{.\EOS\space}%
\providecommand \EOS [0]{\spacefactor3000\relax}%
\providecommand \BibitemShut  [1]{\csname bibitem#1\endcsname}%
\let\auto@bib@innerbib\@empty
\bibitem [{\citenamefont {Daughton}, \citenamefont {Scudder},\ and\
  \citenamefont {Karimabadi}(2006)}]{Daughton2006fully}%
  \BibitemOpen
  \bibfield  {author} {\bibinfo {author} {\bibfnamefont {W.}~\bibnamefont
  {Daughton}}, \bibinfo {author} {\bibfnamefont {J.}~\bibnamefont {Scudder}}, \
  and\ \bibinfo {author} {\bibfnamefont {H.}~\bibnamefont {Karimabadi}},\
  }\bibfield  {title} {\enquote {\bibinfo {title} {Fully kinetic simulations of
  undriven magnetic reconnection with open boundary conditions},}\ }\href@noop
  {} {\bibfield  {journal} {\bibinfo  {journal} {Physics of Plasmas}\ }\textbf
  {\bibinfo {volume} {13}},\ \bibinfo {pages} {072101} (\bibinfo {year}
  {2006})}\BibitemShut {NoStop}%
\bibitem [{\citenamefont {Qiang}\ \emph {et~al.}(2006)\citenamefont {Qiang},
  \citenamefont {Lidia}, \citenamefont {Ryne},\ and\ \citenamefont
  {Limborg-Deprey}}]{Qiang2006Quasistatic}%
  \BibitemOpen
  \bibfield  {author} {\bibinfo {author} {\bibfnamefont {J.}~\bibnamefont
  {Qiang}}, \bibinfo {author} {\bibfnamefont {S.}~\bibnamefont {Lidia}},
  \bibinfo {author} {\bibfnamefont {R.~D.}\ \bibnamefont {Ryne}}, \ and\
  \bibinfo {author} {\bibfnamefont {C.}~\bibnamefont {Limborg-Deprey}},\
  }\bibfield  {title} {\enquote {\bibinfo {title} {Three-dimensional
  quasistatic model for high brightness beam dynamics simulation},}\ }\href
  {\doibase 10.1103/PhysRevSTAB.9.044204} {\bibfield  {journal} {\bibinfo
  {journal} {Phys. Rev. ST Accel. Beams}\ }\textbf {\bibinfo {volume} {9}},\
  \bibinfo {pages} {044204} (\bibinfo {year} {2006})}\BibitemShut {NoStop}%
\bibitem [{\citenamefont {Yang}\ \emph {et~al.}(2010)\citenamefont {Yang},
  \citenamefont {Adelmann}, \citenamefont {Humbel}, \citenamefont {Seidel},
  \citenamefont {Zhang} \emph {et~al.}}]{Yang2010beam}%
  \BibitemOpen
  \bibfield  {author} {\bibinfo {author} {\bibfnamefont {J.}~\bibnamefont
  {Yang}}, \bibinfo {author} {\bibfnamefont {A.}~\bibnamefont {Adelmann}},
  \bibinfo {author} {\bibfnamefont {M.}~\bibnamefont {Humbel}}, \bibinfo
  {author} {\bibfnamefont {M.}~\bibnamefont {Seidel}}, \bibinfo {author}
  {\bibfnamefont {T.}~\bibnamefont {Zhang}},  \emph {et~al.},\ }\bibfield
  {title} {\enquote {\bibinfo {title} {Beam dynamics in high intensity
  cyclotrons including neighboring bunch effects: Model, implementation, and
  application},}\ }\href@noop {} {\bibfield  {journal} {\bibinfo  {journal}
  {Physical Review Special Topics-Accelerators and Beams}\ }\textbf {\bibinfo
  {volume} {13}},\ \bibinfo {pages} {064201} (\bibinfo {year}
  {2010})}\BibitemShut {NoStop}%
\bibitem [{\citenamefont {Cerfon}\ \emph {et~al.}(2013)\citenamefont {Cerfon},
  \citenamefont {Freidberg}, \citenamefont {Parra},\ and\ \citenamefont
  {Antaya}}]{CerfonPRSTAB2013}%
  \BibitemOpen
  \bibfield  {author} {\bibinfo {author} {\bibfnamefont {A.~J.}\ \bibnamefont
  {Cerfon}}, \bibinfo {author} {\bibfnamefont {J.~P.}\ \bibnamefont
  {Freidberg}}, \bibinfo {author} {\bibfnamefont {F.~I.}\ \bibnamefont
  {Parra}}, \ and\ \bibinfo {author} {\bibfnamefont {T.~A.}\ \bibnamefont
  {Antaya}},\ }\bibfield  {title} {\enquote {\bibinfo {title} {Analytic fluid
  theory of beam spiraling in high-intensity cyclotrons},}\ }\href {\doibase
  10.1103/PhysRevSTAB.16.024202} {\bibfield  {journal} {\bibinfo  {journal}
  {Phys. Rev. ST Accel. Beams}\ }\textbf {\bibinfo {volume} {16}},\ \bibinfo
  {pages} {024202} (\bibinfo {year} {2013})}\BibitemShut {NoStop}%
\bibitem [{\citenamefont {Cerfon}(2016)}]{CerfonPRL2016}%
  \BibitemOpen
  \bibfield  {author} {\bibinfo {author} {\bibfnamefont {A.~J.}\ \bibnamefont
  {Cerfon}},\ }\bibfield  {title} {\enquote {\bibinfo {title} {Vortex dynamics
  and shear-layer instability in high-intensity cyclotrons},}\ }\href {\doibase
  10.1103/PhysRevLett.116.174801} {\bibfield  {journal} {\bibinfo  {journal}
  {Phys. Rev. Lett.}\ }\textbf {\bibinfo {volume} {116}},\ \bibinfo {pages}
  {174801} (\bibinfo {year} {2016})}\BibitemShut {NoStop}%
\bibitem [{\citenamefont {Guadagni}\ and\ \citenamefont
  {Cerfon}(2017)}]{Guadagni2017fast}%
  \BibitemOpen
  \bibfield  {author} {\bibinfo {author} {\bibfnamefont {J.}~\bibnamefont
  {Guadagni}}\ and\ \bibinfo {author} {\bibfnamefont {A.~J.}\ \bibnamefont
  {Cerfon}},\ }\bibfield  {title} {\enquote {\bibinfo {title} {Fast and
  spectrally accurate evaluation of gyroaverages in non-periodic
  gyrokinetic-{P}oisson simulations},}\ }\href@noop {} {\bibfield  {journal}
  {\bibinfo  {journal} {Journal of Plasma Physics}\ }\textbf {\bibinfo {volume}
  {83}} (\bibinfo {year} {2017})}\BibitemShut {NoStop}%
\bibitem [{\citenamefont {Askham}\ and\ \citenamefont
  {Cerfon}(2017)}]{Askham2017adaptive}%
  \BibitemOpen
  \bibfield  {author} {\bibinfo {author} {\bibfnamefont {T.}~\bibnamefont
  {Askham}}\ and\ \bibinfo {author} {\bibfnamefont {A.~J.}\ \bibnamefont
  {Cerfon}},\ }\bibfield  {title} {\enquote {\bibinfo {title} {An adaptive fast
  multipole accelerated {P}oisson solver for complex geometries},}\ }\href@noop
  {} {\bibfield  {journal} {\bibinfo  {journal} {Journal of Computational
  Physics}\ }\textbf {\bibinfo {volume} {344}},\ \bibinfo {pages} {1--22}
  (\bibinfo {year} {2017})}\BibitemShut {NoStop}%
\bibitem [{\citenamefont {Fryklund}, \citenamefont {Lehto},\ and\ \citenamefont
  {Tornberg}(2018)}]{Fryklund2018partition}%
  \BibitemOpen
  \bibfield  {author} {\bibinfo {author} {\bibfnamefont {F.}~\bibnamefont
  {Fryklund}}, \bibinfo {author} {\bibfnamefont {E.}~\bibnamefont {Lehto}}, \
  and\ \bibinfo {author} {\bibfnamefont {A.-K.}\ \bibnamefont {Tornberg}},\
  }\bibfield  {title} {\enquote {\bibinfo {title} {Partition of unity extension
  of functions on complex domains},}\ }\href@noop {} {\bibfield  {journal}
  {\bibinfo  {journal} {Journal of Computational Physics}\ }\textbf {\bibinfo
  {volume} {375}},\ \bibinfo {pages} {57--79} (\bibinfo {year}
  {2018})}\BibitemShut {NoStop}%
\bibitem [{\citenamefont {Ethridge}\ and\ \citenamefont
  {Greengard}(2001)}]{Ethridge2001}%
  \BibitemOpen
  \bibfield  {author} {\bibinfo {author} {\bibfnamefont {F.}~\bibnamefont
  {Ethridge}}\ and\ \bibinfo {author} {\bibfnamefont {L.}~\bibnamefont
  {Greengard}},\ }\bibfield  {title} {\enquote {\bibinfo {title} {A new
  fast-multipole accelerated {P}oisson solver in two dimensions},}\ }\href
  {\doibase 10.1137/S1064827500369967} {\bibfield  {journal} {\bibinfo
  {journal} {SIAM Journal on Scientific Computing}\ }\textbf {\bibinfo {volume}
  {23}},\ \bibinfo {pages} {741--760} (\bibinfo {year} {2001})},\ \Eprint
  {http://arxiv.org/abs/https://doi.org/10.1137/S1064827500369967}
  {https://doi.org/10.1137/S1064827500369967} \BibitemShut {NoStop}%
\bibitem [{\citenamefont {McCorquodale}\ \emph {et~al.}(2005)\citenamefont
  {McCorquodale}, \citenamefont {Colella}, \citenamefont {Balls},\ and\
  \citenamefont {Baden}}]{Mccorquodale2005scalable}%
  \BibitemOpen
  \bibfield  {author} {\bibinfo {author} {\bibfnamefont {P.}~\bibnamefont
  {McCorquodale}}, \bibinfo {author} {\bibfnamefont {P.}~\bibnamefont
  {Colella}}, \bibinfo {author} {\bibfnamefont {G.~T.}\ \bibnamefont {Balls}},
  \ and\ \bibinfo {author} {\bibfnamefont {S.~B.}\ \bibnamefont {Baden}},\
  }\bibfield  {title} {\enquote {\bibinfo {title} {A scalable parallel
  {P}oisson solver in three dimensions with infinite-domain boundary
  conditions},}\ }in\ \href@noop {} {\emph {\bibinfo {booktitle} {2005
  International Conference on Parallel Processing Workshops (ICPPW'05)}}}\
  (\bibinfo {organization} {IEEE},\ \bibinfo {year} {2005})\ pp.\ \bibinfo
  {pages} {163--172}\BibitemShut {NoStop}%
\bibitem [{\citenamefont {Langston}, \citenamefont {Greengard},\ and\
  \citenamefont {Zorin}(2011)}]{langston2011}%
  \BibitemOpen
  \bibfield  {author} {\bibinfo {author} {\bibfnamefont {H.}~\bibnamefont
  {Langston}}, \bibinfo {author} {\bibfnamefont {L.}~\bibnamefont {Greengard}},
  \ and\ \bibinfo {author} {\bibfnamefont {D.}~\bibnamefont {Zorin}},\
  }\bibfield  {title} {\enquote {\bibinfo {title} {A free-space adaptive
  {FMM}-based {PDE} solver in three dimensions},}\ }\href@noop {} {\bibfield
  {journal} {\bibinfo  {journal} {Communications in Applied Mathematics and
  Computational Science}\ }\textbf {\bibinfo {volume} {6}},\ \bibinfo {pages}
  {79--122} (\bibinfo {year} {2011})}\BibitemShut {NoStop}%
\bibitem [{\citenamefont {Malhotra}\ and\ \citenamefont
  {Biros}(2015)}]{Malhotra2015pvfmm}%
  \BibitemOpen
  \bibfield  {author} {\bibinfo {author} {\bibfnamefont {D.}~\bibnamefont
  {Malhotra}}\ and\ \bibinfo {author} {\bibfnamefont {G.}~\bibnamefont
  {Biros}},\ }\bibfield  {title} {\enquote {\bibinfo {title} {{PVFMM}: A
  parallel kernel independent {FMM} for particle and volume potentials},}\
  }\href@noop {} {\bibfield  {journal} {\bibinfo  {journal} {Communications in
  Computational Physics}\ }\textbf {\bibinfo {volume} {18}},\ \bibinfo {pages}
  {808--830} (\bibinfo {year} {2015})}\BibitemShut {NoStop}%
\bibitem [{\citenamefont {Hockney}\ and\ \citenamefont
  {Eastwood}(1988)}]{hockney1988computer}%
  \BibitemOpen
  \bibfield  {author} {\bibinfo {author} {\bibfnamefont {R.~W.}\ \bibnamefont
  {Hockney}}\ and\ \bibinfo {author} {\bibfnamefont {J.~W.}\ \bibnamefont
  {Eastwood}},\ }\href@noop {} {\emph {\bibinfo {title} {Computer simulation
  using particles}}}\ (\bibinfo  {publisher} {crc Press},\ \bibinfo {year}
  {1988})\BibitemShut {NoStop}%
\bibitem [{\citenamefont {Budiardja}\ and\ \citenamefont
  {Cardall}(2011)}]{Budiardja2011parallel}%
  \BibitemOpen
  \bibfield  {author} {\bibinfo {author} {\bibfnamefont {R.~D.}\ \bibnamefont
  {Budiardja}}\ and\ \bibinfo {author} {\bibfnamefont {C.~Y.}\ \bibnamefont
  {Cardall}},\ }\bibfield  {title} {\enquote {\bibinfo {title} {Parallel
  {FFT}-based {P}oisson solver for isolated three-dimensional systems},}\
  }\href@noop {} {\bibfield  {journal} {\bibinfo  {journal} {Computer Physics
  Communications}\ }\textbf {\bibinfo {volume} {182}},\ \bibinfo {pages}
  {2265--2275} (\bibinfo {year} {2011})}\BibitemShut {NoStop}%
\bibitem [{\citenamefont {Shishlo}\ \emph {et~al.}(2015)\citenamefont
  {Shishlo}, \citenamefont {Cousineau}, \citenamefont {Holmes},\ and\
  \citenamefont {Gorlov}}]{Shishlo2015particle}%
  \BibitemOpen
  \bibfield  {author} {\bibinfo {author} {\bibfnamefont {A.}~\bibnamefont
  {Shishlo}}, \bibinfo {author} {\bibfnamefont {S.}~\bibnamefont {Cousineau}},
  \bibinfo {author} {\bibfnamefont {J.}~\bibnamefont {Holmes}}, \ and\ \bibinfo
  {author} {\bibfnamefont {T.}~\bibnamefont {Gorlov}},\ }\bibfield  {title}
  {\enquote {\bibinfo {title} {The particle accelerator simulation code
  {PyORBIT}},}\ }\href@noop {} {\bibfield  {journal} {\bibinfo  {journal}
  {Procedia computer science}\ }\textbf {\bibinfo {volume} {51}},\ \bibinfo
  {pages} {1272--1281} (\bibinfo {year} {2015})}\BibitemShut {NoStop}%
\bibitem [{\citenamefont {Adelmann}\ \emph {et~al.}(2019)\citenamefont
  {Adelmann}, \citenamefont {Calvo}, \citenamefont {Frey}, \citenamefont
  {Gsell}, \citenamefont {Locans}, \citenamefont {Metzger-Kraus}, \citenamefont
  {Neveu}, \citenamefont {Rogers}, \citenamefont {Russell}, \citenamefont
  {Sheehy} \emph {et~al.}}]{Adelmann2019opal}%
  \BibitemOpen
  \bibfield  {author} {\bibinfo {author} {\bibfnamefont {A.}~\bibnamefont
  {Adelmann}}, \bibinfo {author} {\bibfnamefont {P.}~\bibnamefont {Calvo}},
  \bibinfo {author} {\bibfnamefont {M.}~\bibnamefont {Frey}}, \bibinfo {author}
  {\bibfnamefont {A.}~\bibnamefont {Gsell}}, \bibinfo {author} {\bibfnamefont
  {U.}~\bibnamefont {Locans}}, \bibinfo {author} {\bibfnamefont
  {C.}~\bibnamefont {Metzger-Kraus}}, \bibinfo {author} {\bibfnamefont
  {N.}~\bibnamefont {Neveu}}, \bibinfo {author} {\bibfnamefont
  {C.}~\bibnamefont {Rogers}}, \bibinfo {author} {\bibfnamefont
  {S.}~\bibnamefont {Russell}}, \bibinfo {author} {\bibfnamefont
  {S.}~\bibnamefont {Sheehy}},  \emph {et~al.},\ }\bibfield  {title} {\enquote
  {\bibinfo {title} {{OPAL} a versatile tool for charged particle accelerator
  simulations},}\ }\href@noop {} {\bibfield  {journal} {\bibinfo  {journal}
  {arXiv preprint arXiv:1905.06654}\ } (\bibinfo {year} {2019})}\BibitemShut
  {NoStop}%
\bibitem [{\citenamefont {Rasmussen}(2011)}]{Rasmussen2011particle}%
  \BibitemOpen
  \bibfield  {author} {\bibinfo {author} {\bibfnamefont {J.~T.}\ \bibnamefont
  {Rasmussen}},\ }\bibfield  {title} {\enquote {\bibinfo {title} {Particle
  methods in bluff body aerodynamics},}\ }\href@noop {} {\bibfield  {journal}
  {\bibinfo  {journal} {Technical University of Denmark}\ } (\bibinfo {year}
  {2011})}\BibitemShut {NoStop}%
\bibitem [{\citenamefont {Chatelain}\ and\ \citenamefont
  {Koumoutsakos}(2010)}]{Chatelain2010}%
  \BibitemOpen
  \bibfield  {author} {\bibinfo {author} {\bibfnamefont {P.}~\bibnamefont
  {Chatelain}}\ and\ \bibinfo {author} {\bibfnamefont {P.}~\bibnamefont
  {Koumoutsakos}},\ }\bibfield  {title} {\enquote {\bibinfo {title} {A
  {F}ourier-based elliptic solver for vortical flows with periodic and
  unbounded directions},}\ }\href {\doibase
  https://doi.org/10.1016/j.jcp.2009.12.035} {\bibfield  {journal} {\bibinfo
  {journal} {Journal of Computational Physics}\ }\textbf {\bibinfo {volume}
  {229}},\ \bibinfo {pages} {2425--2431} (\bibinfo {year} {2010})}\BibitemShut
  {NoStop}%
\bibitem [{\citenamefont {Hejlesen}, \citenamefont {Winckelmans},\ and\
  \citenamefont {Walther}(2019)}]{Hejlesen2019non}%
  \BibitemOpen
  \bibfield  {author} {\bibinfo {author} {\bibfnamefont {M.~M.}\ \bibnamefont
  {Hejlesen}}, \bibinfo {author} {\bibfnamefont {G.}~\bibnamefont
  {Winckelmans}}, \ and\ \bibinfo {author} {\bibfnamefont {J.~H.}\ \bibnamefont
  {Walther}},\ }\bibfield  {title} {\enquote {\bibinfo {title} {Non-singular
  {G}reen’s functions for the unbounded {P}oisson equation in one, two and
  three dimensions},}\ }\href@noop {} {\bibfield  {journal} {\bibinfo
  {journal} {Applied Mathematics Letters}\ }\textbf {\bibinfo {volume} {89}},\
  \bibinfo {pages} {28--34} (\bibinfo {year} {2019})}\BibitemShut {NoStop}%
\bibitem [{\citenamefont {Vico}, \citenamefont {Greengard},\ and\ \citenamefont
  {Ferrando}(2016)}]{Vico2016fast}%
  \BibitemOpen
  \bibfield  {author} {\bibinfo {author} {\bibfnamefont {F.}~\bibnamefont
  {Vico}}, \bibinfo {author} {\bibfnamefont {L.}~\bibnamefont {Greengard}}, \
  and\ \bibinfo {author} {\bibfnamefont {M.}~\bibnamefont {Ferrando}},\
  }\bibfield  {title} {\enquote {\bibinfo {title} {Fast convolution with
  free-space {G}reen's functions},}\ }\href@noop {} {\bibfield  {journal}
  {\bibinfo  {journal} {Journal of Computational Physics}\ }\textbf {\bibinfo
  {volume} {323}},\ \bibinfo {pages} {191--203} (\bibinfo {year}
  {2016})}\BibitemShut {NoStop}%
\bibitem [{\citenamefont {Trefethen}(2000)}]{Trefethen2000spectral}%
  \BibitemOpen
  \bibfield  {author} {\bibinfo {author} {\bibfnamefont {L.~N.}\ \bibnamefont
  {Trefethen}},\ }\href@noop {} {\emph {\bibinfo {title} {Spectral methods in
  {MATLAB}}}}\ (\bibinfo  {publisher} {SIAM},\ \bibinfo {year}
  {2000})\BibitemShut {NoStop}%
\bibitem [{\citenamefont {Trefethen}\ and\ \citenamefont
  {Weideman}(2014)}]{Trefethen2014trapezoidal}%
  \BibitemOpen
  \bibfield  {author} {\bibinfo {author} {\bibfnamefont {L.~N.}\ \bibnamefont
  {Trefethen}}\ and\ \bibinfo {author} {\bibfnamefont {J.}~\bibnamefont
  {Weideman}},\ }\bibfield  {title} {\enquote {\bibinfo {title} {The
  exponentially convergent trapezoidal rule},}\ }\href@noop {} {\bibfield
  {journal} {\bibinfo  {journal} {SIAM Review}\ }\textbf {\bibinfo {volume}
  {56}},\ \bibinfo {pages} {385--458} (\bibinfo {year} {2014})}\BibitemShut
  {NoStop}%
\bibitem [{\citenamefont {Bertschinger}(1998)}]{BertschingerAstro1998}%
  \BibitemOpen
  \bibfield  {author} {\bibinfo {author} {\bibfnamefont {E.}~\bibnamefont
  {Bertschinger}},\ }\bibfield  {title} {\enquote {\bibinfo {title}
  {Simulations of structure formation in the universe},}\ }\href {\doibase
  10.1146/annurev.astro.36.1.599} {\bibfield  {journal} {\bibinfo  {journal}
  {Annual Review of Astronomy and Astrophysics}\ }\textbf {\bibinfo {volume}
  {36}},\ \bibinfo {pages} {599--654} (\bibinfo {year} {1998})},\ \Eprint
  {http://arxiv.org/abs/https://doi.org/10.1146/annurev.astro.36.1.599}
  {https://doi.org/10.1146/annurev.astro.36.1.599} \BibitemShut {NoStop}%
\bibitem [{\citenamefont {Colombi}\ and\ \citenamefont
  {Alard}(2017)}]{Colombi_alard_2017}%
  \BibitemOpen
  \bibfield  {author} {\bibinfo {author} {\bibfnamefont {S.}~\bibnamefont
  {Colombi}}\ and\ \bibinfo {author} {\bibfnamefont {C.}~\bibnamefont
  {Alard}},\ }\bibfield  {title} {\enquote {\bibinfo {title} {A ‘metric’
  semi-lagrangian {V}lasov–{P}oisson solver},}\ }\href {\doibase
  10.1017/S0022377817000393} {\bibfield  {journal} {\bibinfo  {journal}
  {Journal of Plasma Physics}\ }\textbf {\bibinfo {volume} {83}},\ \bibinfo
  {pages} {705830302} (\bibinfo {year} {2017})}\BibitemShut {NoStop}%
\bibitem [{\citenamefont {Tanaka}\ \emph {et~al.}(2017)\citenamefont {Tanaka},
  \citenamefont {Yoshikawa}, \citenamefont {Minoshima},\ and\ \citenamefont
  {Yoshida}}]{Tanaka2017multidimensional}%
  \BibitemOpen
  \bibfield  {author} {\bibinfo {author} {\bibfnamefont {S.}~\bibnamefont
  {Tanaka}}, \bibinfo {author} {\bibfnamefont {K.}~\bibnamefont {Yoshikawa}},
  \bibinfo {author} {\bibfnamefont {T.}~\bibnamefont {Minoshima}}, \ and\
  \bibinfo {author} {\bibfnamefont {N.}~\bibnamefont {Yoshida}},\ }\bibfield
  {title} {\enquote {\bibinfo {title} {Multidimensional {V}lasov--{P}oisson
  simulations with high-order monotonicity-and positivity-preserving
  schemes},}\ }\href@noop {} {\bibfield  {journal} {\bibinfo  {journal} {The
  Astrophysical Journal}\ }\textbf {\bibinfo {volume} {849}},\ \bibinfo {pages}
  {76} (\bibinfo {year} {2017})}\BibitemShut {NoStop}%
\bibitem [{\citenamefont {Passy}\ and\ \citenamefont
  {Bryan}(2014)}]{Passy2014adaptive}%
  \BibitemOpen
  \bibfield  {author} {\bibinfo {author} {\bibfnamefont {J.-C.}\ \bibnamefont
  {Passy}}\ and\ \bibinfo {author} {\bibfnamefont {G.~L.}\ \bibnamefont
  {Bryan}},\ }\bibfield  {title} {\enquote {\bibinfo {title} {An adaptive
  particle-mesh gravity solver for {ENZO}},}\ }\href@noop {} {\bibfield
  {journal} {\bibinfo  {journal} {The Astrophysical Journal Supplement Series}\
  }\textbf {\bibinfo {volume} {215}},\ \bibinfo {pages} {8} (\bibinfo {year}
  {2014})}\BibitemShut {NoStop}%
\bibitem [{\citenamefont {Moon}, \citenamefont {Kim},\ and\ \citenamefont
  {Ostriker}(2019)}]{Moon2019fast}%
  \BibitemOpen
  \bibfield  {author} {\bibinfo {author} {\bibfnamefont {S.}~\bibnamefont
  {Moon}}, \bibinfo {author} {\bibfnamefont {W.-T.}\ \bibnamefont {Kim}}, \
  and\ \bibinfo {author} {\bibfnamefont {E.~C.}\ \bibnamefont {Ostriker}},\
  }\bibfield  {title} {\enquote {\bibinfo {title} {A fast {Poisson} solver of
  second-order accuracy for isolated systems in three-dimensional cartesian and
  cylindrical coordinates},}\ }\href@noop {} {\bibfield  {journal} {\bibinfo
  {journal} {The Astrophysical Journal Supplement Series}\ }\textbf {\bibinfo
  {volume} {241}},\ \bibinfo {pages} {24} (\bibinfo {year} {2019})}\BibitemShut
  {NoStop}%
\bibitem [{\citenamefont {Krasnopolsky}\ \emph {et~al.}(2021)\citenamefont
  {Krasnopolsky}, \citenamefont {Mart{\'{\i}}nez}, \citenamefont {Shang},
  \citenamefont {Tseng},\ and\ \citenamefont {Yen}}]{Krasnopolsky_2021}%
  \BibitemOpen
  \bibfield  {author} {\bibinfo {author} {\bibfnamefont {R.}~\bibnamefont
  {Krasnopolsky}}, \bibinfo {author} {\bibfnamefont {M.~P.}\ \bibnamefont
  {Mart{\'{\i}}nez}}, \bibinfo {author} {\bibfnamefont {H.}~\bibnamefont
  {Shang}}, \bibinfo {author} {\bibfnamefont {Y.-H.}\ \bibnamefont {Tseng}}, \
  and\ \bibinfo {author} {\bibfnamefont {C.-C.}\ \bibnamefont {Yen}},\
  }\bibfield  {title} {\enquote {\bibinfo {title} {Efficient direct method for
  self-gravity in 3d, accelerated by a fast {F}ourier transform},}\ }\href
  {\doibase 10.3847/1538-4365/abca97} {\bibfield  {journal} {\bibinfo
  {journal} {The Astrophysical Journal Supplement Series}\ }\textbf {\bibinfo
  {volume} {252}},\ \bibinfo {pages} {14} (\bibinfo {year} {2021})}\BibitemShut
  {NoStop}%
\end{thebibliography}%

\end{document}